\documentclass[aps,prl,reprint,superscriptaddress,showpacs,longbibliography,floatfix,footnoteinbib]{revtex4-1}
\usepackage[latin9]{inputenc}
\usepackage{color}
\usepackage{array}
\usepackage{float}
\usepackage{bm}
\usepackage{multirow}
\usepackage{amsmath}
\usepackage{amssymb}
\usepackage{graphicx}
\usepackage{setspace}
\usepackage[unicode=true,
 bookmarks=false,
 breaklinks=false,pdfborder={0 0 1},backref=false,colorlinks=true]
 {hyperref}
\hypersetup{pdftitle={Title},
 plainpages=false,pdfpagelabels,linkcolor=blue,urlcolor=blue,citecolor=blue,pdfdisplaydoctitle=true,pdfduplex=DuplexFlipLongEdge}

\makeatletter

\providecommand{\tabularnewline}{\\}

\usepackage{units}\usepackage{wasysym}

\definecolor{orange}{rgb}{0.50, 0.20, 0.0}

\newcommand{\beginsupplement}{%
	\setcounter{page}{1}
	 \renewcommand{\thepage}{SM - \arabic{page}}%
        \setcounter{table}{0}
        \renewcommand{\thetable}{S\arabic{table}}%
        \setcounter{figure}{0}
        \renewcommand{\thefigure}{S\arabic{figure}}%
        \setcounter{section}{0}
        \renewcommand{\thesection}{S\arabic{section}}%
        \setcounter{section}{0}
        \renewcommand{\thesection}{S\arabic{section}}%
        \setcounter{subsection}{0}
        \renewcommand{\thesubsection}{S\arabic{section}.\arabic{subsection}}%
        \setcounter{equation}{0}
        \renewcommand{\theequation}{S\arabic{equation}}%

     }

\usepackage{bm}
\usepackage{braket}

\makeatother

\begin{document}
\noindent\begin{minipage}[t]{1\columnwidth}%
\global\long\def\ket#1{\left| #1\right\rangle }%

\global\long\def\bra#1{\left\langle #1 \right|}%

\global\long\def\kket#1{\left\Vert #1\right\rangle }%

\global\long\def\bbra#1{\left\langle #1\right\Vert }%

\global\long\def\braket#1#2{\left\langle #1\right. \left| #2 \right\rangle }%

\global\long\def\bbrakket#1#2{\left\langle #1\right. \left\Vert #2\right\rangle }%

\global\long\def\av#1{\left\langle #1 \right\rangle }%

\global\long\def\tr{\text{tr}}%

\global\long\def\Tr{\text{Tr}}%

\global\long\def\pd{\partial}%

\global\long\def\im{\text{Im}}%

\global\long\def\re{\text{Re}}%

\global\long\def\sgn{\text{sgn}}%

\global\long\def\Det{\text{Det}}%

\global\long\def\abs#1{\left|#1\right|}%

\global\long\def\up{\uparrow}%

\global\long\def\down{\downarrow}%

\global\long\def\vc#1{\mathbf{#1}}%

\global\long\def\bs#1{\boldsymbol{#1}}%

\global\long\def\t#1{\text{#1}}%
\end{minipage}
\title{Incommensurability-induced sub-ballistic narrow-band-states in twisted
bilayer graphene}
\author{Miguel Gonçalves}
\affiliation{CeFEMA, Instituto Superior Técnico, Universidade de Lisboa, Av. Rovisco
Pais, 1049-001 Lisboa, Portugal}
\author{Hadi Z. Olyaei}
\affiliation{CeFEMA, Instituto Superior Técnico, Universidade de Lisboa, Av. Rovisco
Pais, 1049-001 Lisboa, Portugal}
\author{Bruno Amorim}
\affiliation{Centro de Física das Universidades do Minho e Porto, University of
Minho, Campus of Gualtar, 4710-057, Braga, Portugal}
\author{Rubem Mondaini}
\affiliation{Beijing Computational Science Research Center, Beijing 100193, China}
\author{Pedro Ribeiro}
\affiliation{CeFEMA, Instituto Superior Técnico, Universidade de Lisboa, Av. Rovisco
Pais, 1049-001 Lisboa, Portugal}
\affiliation{Beijing Computational Science Research Center, Beijing 100193, China}
\author{Eduardo V. Castro}
\affiliation{Centro de Física das Universidades do Minho e Porto, Departamento
de Física e Astronomia, Faculdade de Ciências, Universidade do Porto,
4169-007 Porto, Portugal}
\affiliation{Beijing Computational Science Research Center, Beijing 100193, China}
\begin{abstract}
We study the localization properties of electrons in incommensurate
twisted bilayer graphene for small angles, encompassing the narrow-band
regime, by numerically exact means. Sub-ballistic states are found
within the narrow-band region around the magic angle. Such states
are delocalized in momentum-space and follow non-Poissonian level
statistics, in contrast with their ballistic counterparts found for
close commensurate angles. Transport results corroborate this picture:
for large enough systems, the conductance decreases with system size
for incommensurate angles within the sub-ballistic regime. Our results
show that incommensurability/quasiperiodicity effects are of crucial importance in
the narrow-band regime. The incommensurate nature of a general twist
angle must therefore be taken into account for an accurate description
of magic-angle twisted bilayer graphene.
\end{abstract}
\maketitle
\begin{spacing}{0.95}
\noindent Narrow-band electronic systems are natural platforms to
search for exotic physics. Otherwise modest perturbations can here
attain energies comparable to the bandwidth, yielding a non-perturbative
reorganization of the eigenstates into phases of matter often very
different from their parent state. Recently, twisted bilayer graphene
(tBLG) has emerged as a paradigmatic system displaying this mechanism.
When the twist angle between the layers approaches the
so-called \emph{magic angle,} $\theta\approx1.1^{\circ}$, extremely
narrow, nearly flat, bands appear at low energies \cite{PhysRevLett.99.256802,Bistritzer2011a,PhysRevB.86.155449,TramblydeLaissardiere2010a,SuarezMorell2010,Shallcross2010}.
In this regime, superconductivity \cite{Cao2018} and correlated insulating
phases \cite{Cao2018a} were observed around integer electronic
fillings, pointing to relevant electron-electron interactions \cite{Balents2020}.
The similarities with the cuprates \cite{Bednorz:1986tc} sparked
redoubled interest in these effects, which emerge here in a simpler,
cleaner and highly tunable system, where all ingredients can potentially
be easily isolated and understood. This triggered intense theoretical \cite{Kang2018,Zou2018,Po2018a,Tarnopolsky2019,Haddadi2019,Song2019,Carr2018a,
Chen2019a,PhysRevB.100.205113,Po2018b,Yuan2018a,Po2018,Kennes2018,Peltonen2018,Isobe2018,PhysRevB.102.155149,PhysRevB.102.045107}
and experimental \cite{Chung2018,Yankowitz2019,Lu2019,Jiang2019,Sharpe2019,Kerelsky2019,Choi2019,Xie2019a,Tomarken2019,Serlin2019}
research.
\end{spacing}

Due to the high degree of sample-purity, models of tBLG typically
assume a Fermi gas with a flat-band dispersion as a starting point
\cite{Kang2018,Zou2018,Po2018a,Chen2019a,Haule2019,Kennes2018,Peltonen2018,Isobe2018,Yuan2018,Yuan2018a}.
This is a major difference with respect to the cuprates, as disorder
is intrinsic to doped Mott insulators \cite{Pan2001} and has even
been observed to increase the critical temperature \cite{Leroux2019}.

Nonetheless, in Refs.~\cite{PhysRevB.99.165430,Park2018}, localization
has been predicted for the recently observed ``dodecagonal
graphene quasicrystal'' \cite{Yao2018,Ahn2018}, a tBLG with $\theta=30^{\circ}$.
This phenomenon is due to the quasiperiodic nature of the
system for incommensurate values of $\theta$. Incommensurability
was also shown to induce quantum phase transitions in two-dimensional
(2d) models \cite{Fu2018}, including the so-called chiral limit
of tBLG for moderate twist angles ($\theta\simeq9^{\circ}$), yielding
a critical ``magic-angle semimetallic'' state with a multifractal
momentum-space wave function. For more realistic models of tBLG, the
proximity to commensurate angles with small unit cells,
has been shown to imprint sharp conductance signatures \cite{Olyaei2020}.
However, for small angles, $\theta\lesssim3^{\circ}$, these features
seem to be washed away, in accordance with the general belief underlying
continuous models \cite{Tarnopolsky2019}.

Incommensurability can doubtlessly induce localization \cite{PhysRevB.100.144202,PhysRevB.101.014205,PhysRevB.99.054211}
or multifractallity in 2d \cite{Fu2018}. However, at this point,
it is not clear how the commensurate/incommensurate nature of the
tBLG structure affects the properties of eigenstates in the small-angle narrow-band regime. As in tBLG several energy
scales are comparable, understanding the role played by incommensurability
is essential to devise effective interacting models able to faithfully
capture their competition.

In this letter we address the effects of incommensurability/quasiperiodicity in the
nature of the narrow-band states and on the transport properties of
tBLG, neglecting electron-electron interactions. We employ a number
of numerically exact methods to show that incommensurate angles induce
momentum-space delocalization of the narrow-band states, whereas for
commensurate angles eigenstates are ballistic with a localized momentum-space
wave function. The presence of a sub-ballistic regime is corroborated
by finite-size scaling analysis of the conductance which decreases
with system size for incommensurate structures, while saturating
for commensurate ones.

\begin{figure}
\centering{}\includegraphics[width=1\columnwidth]{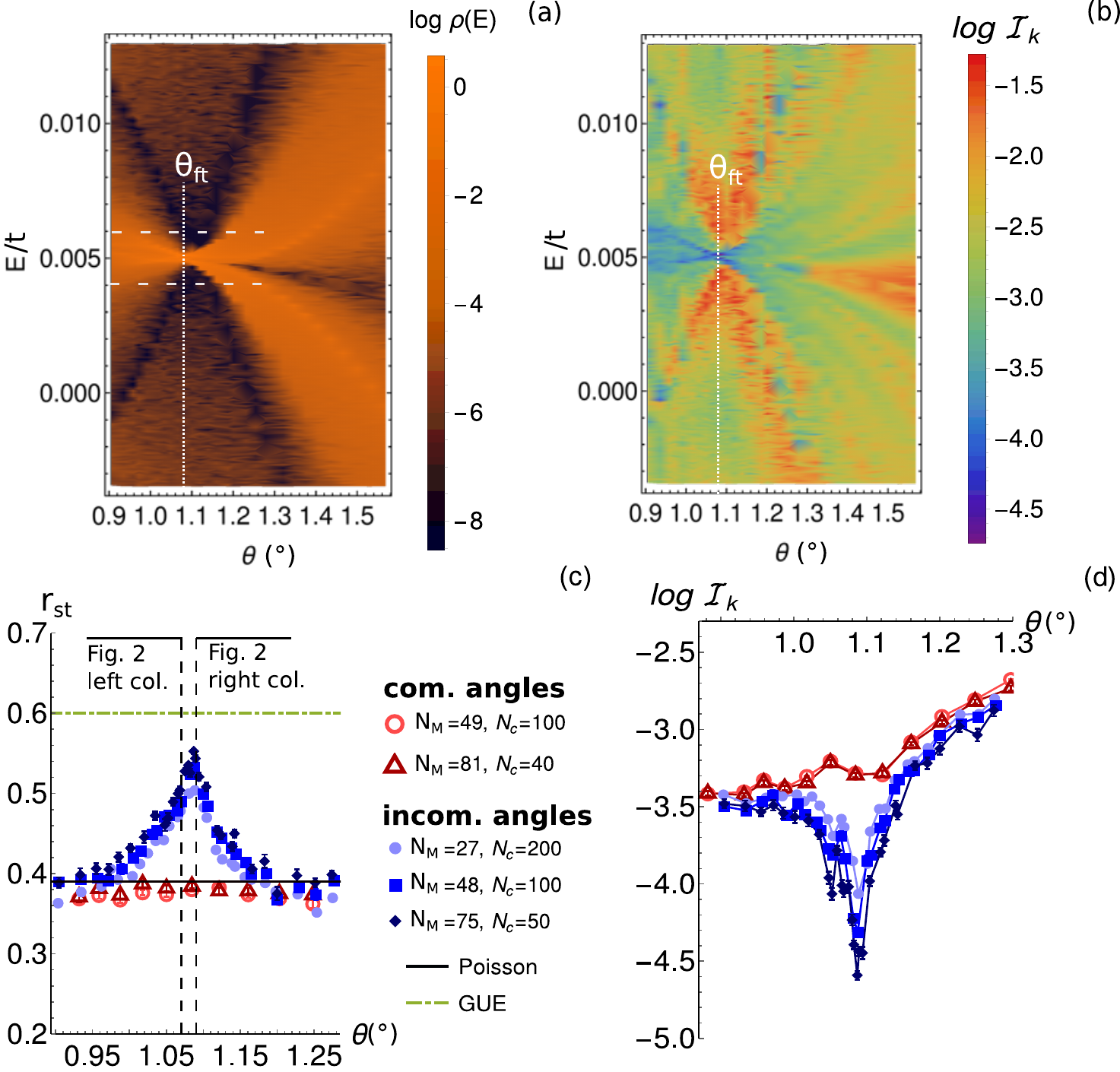}\caption{(a,b) DOS, $\rho$, and momentum-space inverse participation ratio,
$\mathcal{I}_{k}$, for incommensurate structures and variable $\theta$. The system sizes range
between $N=2\times10^{5}-8\times10^{5}$ sites, $N_{{\rm M}}=48$ moiré cells and an average
over $N_{c}=100$ realizations was performed. (c,d) Level spacing,
$r_{{\rm st}}$, and $\mathcal{I}_{k}$ averaged over the energy window
between dashed line in (a), computed both for a set of commensurate angles
(variable $m$, $r=1$, $n=7,9$) and incommensurate angles (variable
$m$, $r=9,12,15$, $n=1$), for system sizes going up
to more than $N=10^{6}$ ($N_{{\rm M}}=81$ and $N_{{\rm M}}=75$).
\label{fig:panel1}}
\end{figure}

\begin{figure}
\centering{}\includegraphics[width=1\columnwidth]{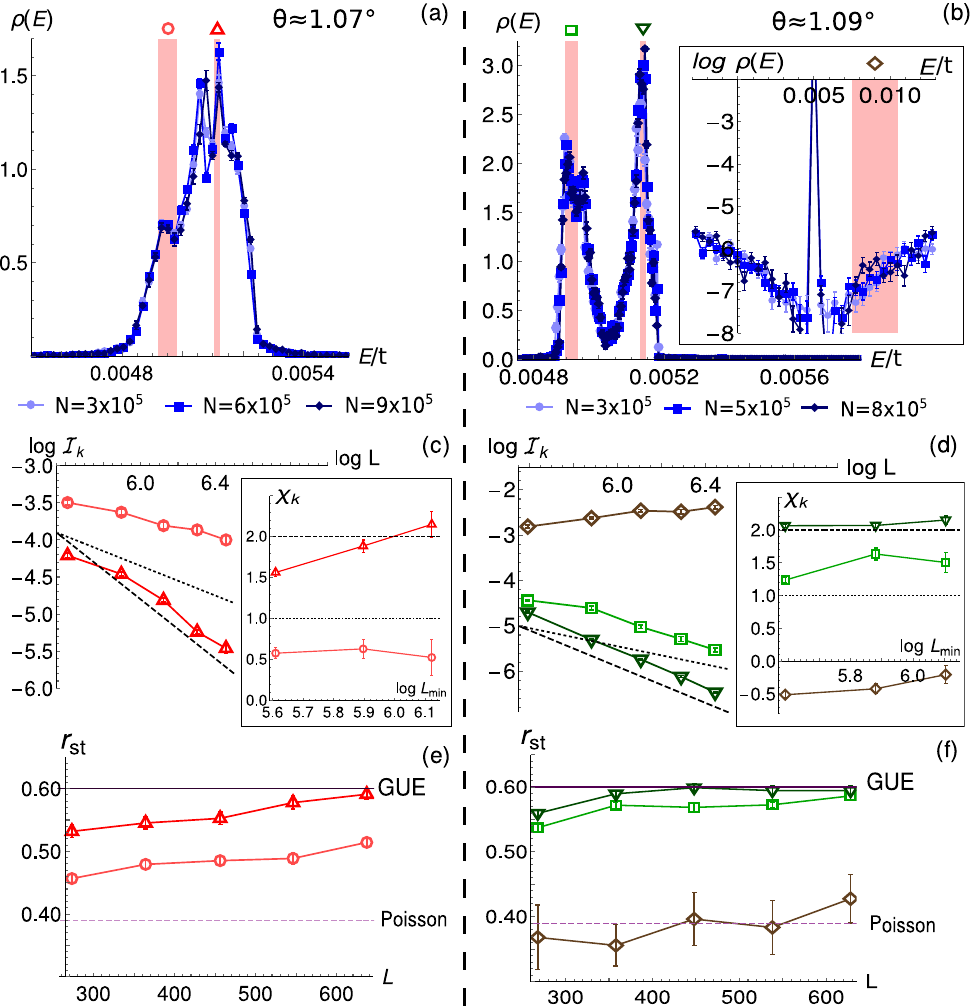}\caption{(a,b) DOS for incommensurate structures of different sizes, for the
angles marked in Fig.$\,$\ref{fig:panel1}(c). The inset in (b)
shows a log-scale plot of $\rho(E)$ for a wider energy range. (c-f)
Average $r_{{\rm st}}$ and $\mathcal{I}_{k}$ for a small energy window around
selected energies, indicated by the shaded areas in (a,b) along with marker on top,
as a function of the linear system size $L\propto N^{1/2}$ (the largest
systems correspond to $N\gtrsim10^{6}$, see \cite{SM}).
The dotted
and dashed lines in (c,d) correspond respectively to $\mathcal{I}_{k}\sim L^{-1},L^{-2}$.
The insets in (c,d) contain the quantity $\chi_{k}=-d\log\mathcal{I}_{k}/d\log L$, computed by fitting the data points $(\log L,\log\mathcal{I}_{k})$,
considering only points with $L\protect\geq L_{{\rm min}}$. Complete
information on the data used in Figs.$\,$(a-f) can be found in table
S1 in \cite{SM}.
\label{fig:non_ballistic_dos_scalings}}
\end{figure}

\paragraph{Model and Methods.---}

We study a model of tBLG with the tight-binding Hamiltonian
\begin{equation}
\begin{aligned} & H=-t\sum_{l,\langle\bm{r}_{l,A},\bm{r}_{l,B}\rangle}c_{l,A}^{\dagger}(\bm{r}_{l,A})c_{l,B}(\bm{r}_{l,B})\\
 & +\sum_{|\bm{r}_{1,\alpha}-\bm{r}_{2,\beta}|<\Lambda}t_{\perp}(|\bm{r}_{1,\alpha}-\bm{r}_{2,\beta}|)c_{1,\alpha}^{\dagger}(\bm{r}_{1,\alpha})c_{2,\beta}(\bm{r}_{2,\beta})+\text{h.c.},
\end{aligned}
\end{equation}
where the first term describes the nearest-neighbor intralayer hopping $t$ \footnote{We also tested models with intralayer hoppings up to third nearest neighbors and observed that the results remained qualitatively unaffected up to the expected shift in the energy of the Dirac points and a slight change in the narrow band angle $\theta_{\text{ft}}$.}, and $l=1,2$ is the layer index. The second term
models the interlayer hopping, with $\bm{r}_{l,\alpha}$ the in-plane
position in layer $l$ and sublattice $\alpha=A,B$. The interlayer
hopping $t_{\perp}(r)$ is parameterized in terms of Slater-Koster
parameters used in Refs.$\,$\cite{DeTramblyLaissardiere2010Original,PhysRevB.86.125413,PhysRevMaterials.2.034004}:
$r_{\triangle}^{2}t_{\perp}(r)=d_{\perp}^{2}V_{pp\sigma}(r_{\triangle})+r^{2}V_{pp\pi}(r_{\triangle})$
where $r_{\triangle}=\sqrt{d_{\perp}^{2}+r^{2}}$, $V_{pp\sigma}(r)=t_{\perp}\exp[(d_{\perp}-r)/\delta]$,
$V_{pp\pi}(r)=-t\exp[(d-r)/\delta]$, $d=0.142\,{\rm nm}$ is the
C-C distance, $d_{\perp}=0.335\,{\rm nm}$ is the distance between
layers and $\delta=0.184a$, with $a=0.246\,{\rm nm}$ the monolayer
lattice constant. We used $t=2.7\,{\rm eV}$ and $t_{\perp}=0.48\,{\rm eV}$.
For the calculations presented hereafter, energy and length scales
are measured in units of $t$ and $a$, respectively.

Interlayer hoppings were considered only among atoms with in-plane
distance $r<\Lambda$. For a small cutoff, $\Lambda\sim d$, the narrow-band angle, $\theta_{\text{ft}}$, is very sensitive
to relative translations of the layers. This effect washes-away for
larger $\Lambda$. To mitigate finite-size effects we set $\Lambda=2.6d$,
above which our numerical results remained unchanged. 

A tBLG lattice with periodic boundary conditions can be defined by
the integers $(m,r,n)$, where $(m,r)$ are coprime numbers that determine
the commensurate twist angle $\theta(m,r)$ \cite{PhysRevB.86.155449},
and $n$ is the linear number of supercells in the system, i.e., the
lattice contains $n^{2}$ supercells. The superlattice basis vectors,
the commensurate twist angle $\theta\left(m,r\right)$ \cite{PhysRevB.86.155449}
, and the number of moiré pattern cells $N_{\text{M}}\left(n,r\right)$
as a function of the integer tuples are given in the supplemental
material (SM) \cite{SM}. 
We follow a well established method that consists of approximating an infinite incommensurate structure (here characterised by twist angle $\theta_{\text{ic}}$) by a sequence of approximants with increasing lattice size (here taken to be $n=1$ structures with $\theta(m,r)\simeq \theta_{\text{ic}})$ \footnote{ This method was originally proposed in Ref.$\,$\citep{PhysRevLett.43.1954}. See also Ref.$\,$\citep{PhysRevB.98.134201} for an example of a recent application}.  
An approximant structure may encompass many moiré cells (for $r_{i}\neq1$), its unit cell coincides with the
size of the system. 
Commensurate structures are obtained by taking
$n>1$ and a finite-size scaling analysis is
performed by increasing $n$ with fixed $(m,r)$. The main advantage
of this method is to eliminate boundary effects, present for open
systems, by always considering closed boundary
conditions \footnote{It is important to advert that the study of momentum-space localization
in finite systems requires closed boundary conditions, which motivates
the construction of approximant structures previously described. For
open boundaries, one-dimensional edge defects are sufficient to induce momentum-space delocalization even for commensurate
structures.}.

To characterize the system's eigenstates near the narrow-band energies,
we used Krylov-Schur exact diagonalization (ED) with shift-invert
\cite{Slepc,Petsc}. To further mitigate finite-size effects, we average
all relevant quantities with respect to a phase-twist we introduce
in the boundary conditions and to a random relative stacking translation,
$\bm{\delta}_{\mathrm{t}}$, between the two layers. We approximate
the density of states (DOS) by binning the energies, $\rho(E_{i})=N_{s}(E_{i})/\left(\Delta EN\right)$,
where $N_{s}(E_{i})$ is the number of states inside the bin centered
around energy $E_{i}$ and width $\Delta E$, and $N$ is the total
number of sites. We also study the eigenstates' momentum-space localization
properties through the momentum-space inverse participation ratio
(${\rm IPR_{k}}$) \cite{Wobst2002}, 
\begin{equation}
\mathcal{I}_{k}=\Bigl(\sum_{\bm{k},\alpha}|\Psi_{l,\bm{k},\alpha}|^{2}\Bigr)^{-2}\sum_{\bm{k},\alpha}|\Psi_{l,\bm{k},\alpha}|^{4},\label{eq:IPRkq-1}
\end{equation}

\noindent where $\Psi_{l,\bm{k},\alpha}$ is the eigenstate amplitude
in layer $l$ momentum $\boldsymbol{k}$ and sublattice $\alpha$.
For wavefunctions localized in momentum-space, $\mathcal{I}_{k}\sim L^{0}$,
where $L$ is the linear system size \footnote{In fact, $L$ is the linear size of the rhombus used to compute $\mathcal{I}_{k}$,
which is proportional to $N^{1/2}$: this size is what sets the resolution
for the computation of $\mathcal{I}_{k}$, see SM \cite{SM} for details.}, whereas a $\mathcal{I}_{k}\sim L^{-\nu}$, with $\nu>0,$ indicates
momentum-space delocalization. We checked that the real-space inverse
participation ratio always scales with $L^{-2}$, signaling no real-space
localization. Therefore, $\mathcal{I}_{k}\sim L^{-\nu}$,
with $\nu>0$, indicates the presence of sub-ballistic states.

To complement the eigenstate's analysis, we study the statistics of
the energy levels through the quantity $r_{{\rm st}}=\langle r_{i}\rangle_{E\in E_{w},N_{c}}$,
where $r_{i}$ is defined as $r_{i}=\min(\Delta E_{i},\Delta E_{i+1})/\max(\Delta E_{i},\Delta E_{i+1}),$
with $\Delta E_{i}=E_{i}-E_{i-1}$ (for ordered levels $E_{i}>E_{i-1}$)
and where the average, $\langle...\rangle_{E\in E_{w},N_{c}}$, is
first performed over all the eigenvalues within an energy window $E_{w}$
for each realization and then over $N_{c}$ realizations of boundary
twists and stacking shifts. The relevant known values \cite{Atas2013}
for $r_{{\rm st}}$ are: (i) $r_{{\rm st}}^{{\rm Poisson}}\approx0.39$
if the spacings follow a Poisson distribution; (ii) $r_{{\rm st}}^{{\rm GUE}}\approx0.6$
for the Gaussian unitary ensemble (GUE), when the Hamiltonian breaks
time-reversal symmetry (here due to twisted boundary conditions). Case (i) applies when the wavefunction is
localized in some basis: this includes ballistic (momentum-space localized)
or insulating (real-space localized) states. Case (ii) applies when
the wavefunction is delocalized in any basis.

Finally, we analyze transport properties by computing the conductance,
$G$. We considered two semi-infinite stripes of single-layer graphene
that overlap within a region of fixed width and variable length, $\sim L$.
The graphene leads are rendered metallic by doping (see SM \cite{SM}
for more details).

\paragraph{Commensurate vs. incommensurate.---}

We start by providing a general overview on the differences between
commensurate and incommensurate structures around the first magic-angle. Fig.$\,$\ref{fig:panel1}(a,b),
depict the DOS and $\mathcal{I}_{k}$ of incommensurate structures
for different energies and angles with a fixed number of moiré cells $N_{\t M}=48$. As expected, a narrow-band occurs for the \emph{magic-angle}
$\theta_{\text{ft}}\approx1.09^{\circ}$, close to the merging of
two van-Hove singularities (VHS) present at larger $\theta$. Remarkably,
Fig.$\,$\ref{fig:panel1}(b) shows that $\mathcal{I}_{k}$ becomes
very small for energies within the narrow-band. This is not observed
for commensurate structures, as justified below.

Figs.$\,$\ref{fig:panel1}(c,d) show $r_{{\rm st}}$ and $\mathcal{I}_{k}$
averaged within the energy window represented by the horizontal dashed
lines in Fig.$\,$\ref{fig:panel1}(a), both for incommensurate ($r=9,12,15$;
$n=1$) and commensurate ($r=1$; $n=7,9$) structures. As expected
for ballistic states, for commensurate angles, $r_{{\rm st}}$ follows
Poisson statistics and $\mathcal{I}_{k}$ is independent of the system
size. Conversely, for incommensurate angles: (i) $r_{{\rm st}}$ raises
above the Poisson value within a finite interval of (incommensurate)
angles, reaching a maximum value for $\theta=\theta_{{\rm ft}}$;
(ii) For the same interval of angles, $\mathcal{I}_{k}$ scales down
with system size, reaching a minimum for $\theta=\theta_{\text{ft}}$.
Outside this regime, $\mathcal{I}_{k}$ becomes $L$-independent and
approaches the value obtained for commensurate structures {[}Fig.$\,$\ref{fig:panel1}(d){]}.

These results suggest that, as a function of $\theta$, there is a
``collision'' of energy bands around $\theta_{\text{ft}}$, with
subsequent band inversion. We confirm such phenomena when the narrow
band regime is tuned by changing the amplitude of $t_{\perp}$ (see
\cite{SM}). It is plausible that the same happens when changing $\theta$
(for fixed $t_{\perp}$), but a similar analysis is prevented within
our setup by changes in system size with $\theta$. Within this picture,
the narrow-band regime corresponds to the collision area where states
from the two bands become highly mixed (as seen by $\mathcal{I}_{k}$)
and their energy levels repel (as seen by $r_{{\rm st}}$). Moreover,
at least a finite portion of the states inside the considered energy
window have sub-ballistic properties within a finite interval of angles
that we now explore in more detail.

\begin{figure}
\centering{}\includegraphics[width=1\columnwidth]{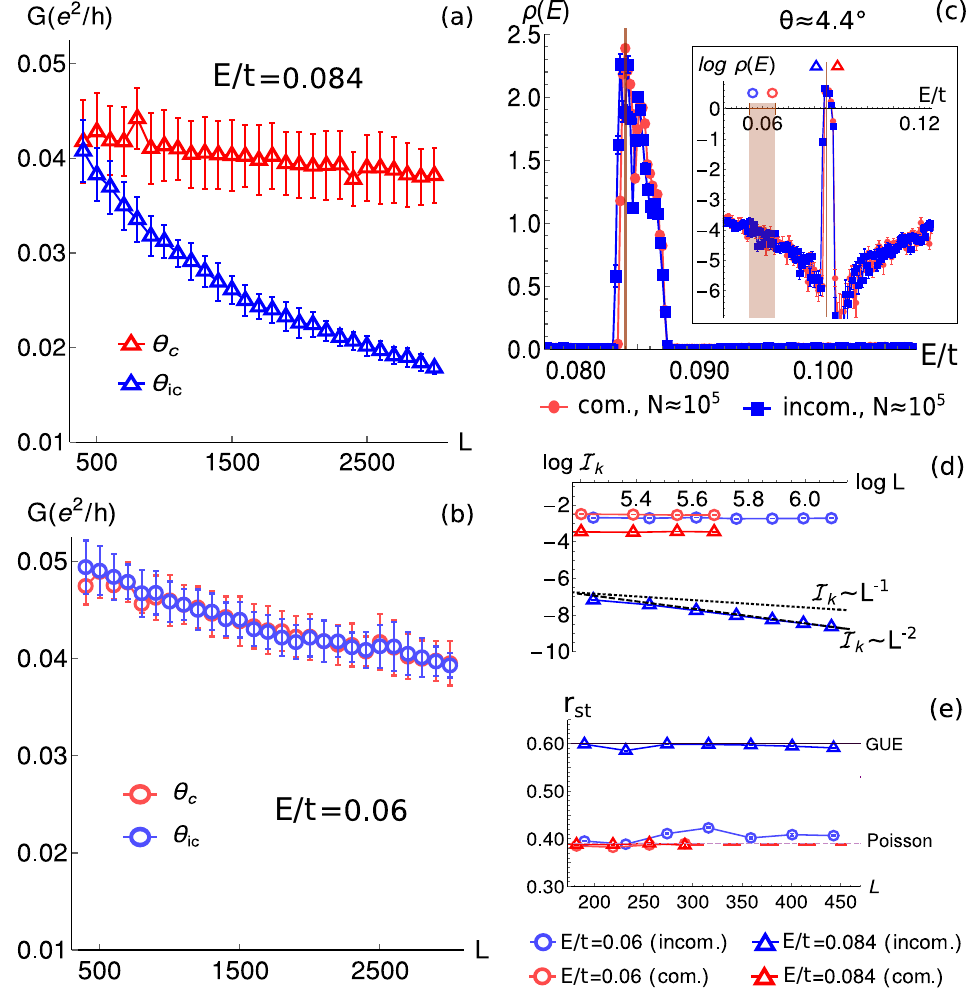}\caption{
(a,b) Finite-size scaling analysis of the conductance for commensurate,
$\theta_{{\rm c}}$, and incommensurate, $\theta_{{\rm ic}}$, angles
near $\theta_{\text{ft}}$: (a) is for $E=0.084$, inside
the narrow band; while$\,$(b) is for $E=0.06$, outside. The error bars correspond to the error of the average conductance computed for 25 different stackings between the graphene layers (see \cite{SM}). \textbf{
}(c) DOS for $\theta_{{\rm c}}$ and $\theta_{{\rm ic}}$. Inset:
log-scale plot for a wider energy range. (d,e) Finite-size analysis
of $\mathcal{I}_{k}$ and $r_{{\rm st}}$, averaged on small energy
windows {[}shaded areas in (c) around $E=0.06$ and $E=0.084$,
with widths $\Delta E=10^{-2}$ and $2\times10^{-4}${]} for $\theta_{c}$,
and $\theta_{{\rm ic}}$ (see table S2
in \cite{SM}). For $E=0.06$, the DOS is small and
a larger energy window is used to increase statistics. Plotmarkers
are used in the scaling analysis of (d,e). Dotted and dashed lines
in (d) correspond respectively to $\mathcal{I}_{k}\sim L^{-1},L^{-2}$.
\label{fig:conductance_vs_ED}}
\end{figure}

\paragraph{Sub-ballistic regime.---}

We provide an energy resolved
and finite-size scaling analysis. Figure $\,$\ref{fig:non_ballistic_dos_scalings}
shows two angles chosen in order to observe merged VHSs in \ref{fig:non_ballistic_dos_scalings}(a),
and a minimal bandwidth in \ref{fig:non_ballistic_dos_scalings}(b).

Figure$\,$\ref{fig:non_ballistic_dos_scalings}(c,d) and~\ref{fig:non_ballistic_dos_scalings}(e,f)
depict, respectively, $\mathcal{I}_{k}$ and $r_{{\rm st}}$ averaged
over the shaded energy windows of Figs.$\,$\ref{fig:non_ballistic_dos_scalings}(a,b).
The approximant series was chosen to ensure a monotonic approach to
the desired incommensurate angle upon increasing system size (see
\cite{SM}), and we ensure an overall angle variation in the series
below $10''$. States with different scaling behaviors arise at different
energies. Typically, when $\rho(E)$ is larger, the $\mathcal{I}_{k}$
is smaller and decreases faster with system size. The latter correlates
with a faster scaling of $r_{{\rm st}}$ towards the GUE value. For
instance, for energies corresponding to the larger DOS observed in
Fig.$\,$\ref{fig:non_ballistic_dos_scalings}(b), $\mathcal{I}_{k}$
reaches its smallest value, scaling as $\mathcal{I}_{k}\sim L^{-2}$,
and $r_{{\rm st}}$ converges very quickly with $L$ into the GUE
value. For other energies within the narrow-band, we generically have
$0<\chi_{k}<2$, where $\chi_{k}=-d\log\mathcal{I}_{k}/d\log L$,
and $r_{{\rm st}}^{{\rm Poisson}}<r_{{\rm st}}<r_{{\rm st}}^{{\rm GUE}}$.
Even if there are considerable finite-size effects, as $\chi_{k}$
and $r_{{\rm st}}$ still seem to be increasing with $L$ up to the
sizes we can reach (see Fig.$\,$\ref{fig:non_ballistic_dos_scalings}),
it is clear that states in this region are sub-ballistic. These findings
are further corroborated by the transport results of the next section.

Outside the narrow-band region, states are ballistic ($r_{{\rm st}}=r_{{\rm st}}^{{\rm Poisson}}$
and $\mathcal{I}_{k}\sim L^{0}$) even in regions with a highly suppressed
DOS. An example is given in Fig.$\,$\ref{fig:non_ballistic_dos_scalings}(f)
for the states within the shaded area in the inset of Fig.$\,$\ref{fig:non_ballistic_dos_scalings}(b).

Finally, for angles above and below the narrow-band regime in Fig.$\,$\ref{fig:non_ballistic_dos_scalings},
but still within the sub-ballistic range, states exhibiting non-ballistic properties are also observed. However, their limiting behavior is less conclusive (see \cite{SM}).

\paragraph{Conductance.---}

In order to understand the consequences of the sub-ballistic states
in transport, we computed the conductance, $G$. Unfortunately, for the parameters above, finite-size
effects are too severe to draw systematic conclusions (see SM \cite{SM}),
as the largest systems attainable contain only $N_{\t M}\sim10^{2}$
moiré cells. To mitigate finite-size effects by increasing
the number of moiré cells simulated, we increased the interlayer coupling
to $t_{\perp}=1.9\,{\rm eV}$, a well-known procedure for shifting
the narrow-band regime to larger angles ($\theta_{\text{ft}}\sim4.4^{\circ}$,
in this case), with smaller period moiré patterns \cite{PhysRevLett.99.256802,Bistritzer2011a,PhysRevB.86.155449}.
Representative conductance results for $t_{\perp}=1.9\,{\rm eV}$
are given in Figs.$\,$\ref{fig:conductance_vs_ED}(a,b), together
with an ED analysis in Figs.$\,$\ref{fig:conductance_vs_ED}(c-e).
In Fig.$\,$\ref{fig:conductance_vs_ED}(a), $G(L)$ is computed for
an energy within the narrow-band. In the commensurate case, $G(L)$
converges with $L$ as expected. Conversely, it decreases with $L$
in the incommensurate case, hinting sub-ballistic transport. For smaller $L$, the conductance is very similar in both cases,
showing that large enough systems are crucial to observe incommensurability
effects in the conductance. On the other hand, ED captures a sub-ballistic
behavior even for the smaller systems used: a small, $L$-decreasing
$\mathcal{I}_{k}$ and GUE statistics are observed in Figs.$\,$\ref{fig:conductance_vs_ED}(d,e).

For energies outside the narrow-band, where the DOS is very small
{[}see inset in Fig.$\,$\ref{fig:conductance_vs_ED}(c){]}, the ED
and conductance results are very similar for commensurate and incommensurate
angles {[}Figs.$\,$\ref{fig:conductance_vs_ED}(b,d){]}, suggesting
that incommensurability effects become unimportant.

\paragraph{Discussion.---}

We studied the eigenstate localization and transport properties of
tBLG for commensurate and incommensurate angles around the first
magic angle. For a finite interval of incommensurate angles, encompassing
the narrow-band regime, eigenstate delocalization in momentum-space
is concomitant with non-Poisson energy level statistics and with a
decrease of the conductance with system size. This is compatible with
sub-ballistic transport and contrasts with commensurate angles, or
with angles away from the narrow-band region, where transport is
ballistic. Moreover, the scaling of $\mathcal{I}_{k}$
with system size seem to indicate diffusive behavior, although the
results from conductance scaling are unclear with the available system
sizes. The present results indicate that the scenario of a ``magic
angle semimetal'' with momentum-space delocalized wave functions,
proposed in Ref.~\cite{Fu2018} for a model of moderate angle chiral
tBLG, extends to models of incommensurate tBLG with twist
angles close to the experimentally relevant magic angle $\theta \approx1.1^{\circ}$.

Our results have major implications for the low energy properties
of tBLG as, in the narrow-band regime, incommensurability alone breaks down a Bloch wave description even for perfectly
clean samples. In particular, an analysis of the influence of correlations should equally account for the incommensurate nature of tBLG. These effects have been overlooked in the vast majority of theoretical studies on tBLG, namely the ones starting from continuum models \citep{PhysRevLett.99.256802,Bistritzer2011a,PhysRevB.86.155449}.
Our findings may also be relevant to the enhanced,
linear temperature resistivity observed in tBLG at the magic angle
\cite{Polshyn2019,Cao2020}. Even though we restricted the study to
small angle tBLG, we anticipate that the present results also apply
to other systems with concomitant narrow-bands and incomensurability:
tBLG at large angles \cite{Pal2019}, double bilayers \cite{Shen2019,Liu2019,Cao2019},
twisted bilayers of transition metal dichalcogenides \cite{Wu2019,Wang2019}. We also note that even though our model does not account for lattice relaxation, recent experimental results showed that near the magic-angle, both the relaxed and unrelaxed structures are stable, being possible to change between them by applying a STM tip pulse \citep{PhysRevLett.125.236102}. Moreover, even in the relaxed case, the narrow-band is still
present \citep{PhysRevB.99.205134} and may even be narrower \citep{PhysRevB.96.075311} with incommensurability effects possibly enhanced. 

As single-particle localization can give rise to many-body localization once interactions are included \citep{PhysRevB.87.134202,Schreiber842}, we expect the single-particle properties here reported to also play an important role once interactions are considered. 

Finally, we checked that our results are robust to relatively strong Anderson-like disorder (of the order of the narrow-band's width, see \cite{SM}). Interestingly, commensurate structures are more fragile and, at stronger disorder, their properties approach those of disorder-free incommensurate structures. 



\begin{acknowledgments}
MG, HO and PR acknowledge partial support from Fundação para a Ciência
e Tecnologia (Portugal) through Grant and UID/CTM/04540/2019. BA and
EVC acknowledge support from FCT-Portugal through Grant No.~UIDB/04650/2020.
MG acknowledges further support from FCT-Portugal through the Grant
SFRH/BD/145152/2019. HO acknowledges further support through the Grant
PD/BD/113649/2015. BA acknowledges further support from FCT-Portugal
through Grant No. CEECIND/02936/2017. RM acknowledges support from
NSFC Grants No. 11674021, No. 11851110757, No. 11974039, and NSAF-U1930402.
The hospitality of the Computational Science Research Center, Beijing,
China, where this work was initiated, is also acknowledged. We finally acknowledge the Tianhe-2JK cluster at the Beijing Computational
Science Research Center (CSRC), the Baltasar-Sete-Sóis cluster,
supported by V. Cardoso\textquoteright s H2020 ERC Consolidator Grant
no. MaGRaTh-646597, and the OBLIVION supercomputer (based at the High Performance Computing Center - University of Évora) funded by the ENGAGE SKA Research Infrastructure (reference POCI-01-0145-FEDER-022217 - COMPETE 2020 and the Foundation for Science and Technology, Portugal) and by the BigData@UE project (reference ALT20-03-0246-FEDER-000033 - FEDER and the Alentejo 2020 Regional Operational Program. Computer assistance was provided by CSRC, CENTRA/IST and the OBLIVION support team.
\end{acknowledgments}

\bibliography{tBLG_arxiv_v2}

\clearpage \onecolumngrid

\beginsupplement
\begin{center}
\textbf{\large{}Supplemental Material for: }\\
\textbf{\large{} \vspace{0.1cm}
Incommensurability-induced sub-ballistic narrow-band-states in twisted bilayer
graphene}{\large\par}
\par\end{center}

\vspace{0.3cm}

\begin{center}
\vspace{0.3cm}
\par\end{center}

\vspace{0.6cm}

\twocolumngrid

\tableofcontents{}

\section{Finite-size scaling method for exact diagonalization}

As mentioned in the main text, in order to make a finite-size scaling
analysis with our method, we need to slightly change the angle to
work with incommensurate structures only. Here, we discuss this procedure
in more detail.

We build structures characterized by the integers $(m,r,n)$ ($m$
and $r$ being two coprime integers), with twist angles $\theta=\theta(m,r)$
given by

\begin{equation}
\cos\theta=\frac{3m^{2}+3mr+r^{2}/2}{3m^{2}+3mr+r^{2}},\label{eq:cos_theta}
\end{equation}
number of sites $N=N(m,r,n)$ and number of moiré patterns, $N_{M}=N_{M}(n,r)$:

\begin{equation}
N_{M}=\begin{cases}
n^{2}r^{2} & ,\mod(r,3)\neq0\\
n^{2}r^{2}/3 & ,\textrm{mod}(r,3)=0
\end{cases}.
\end{equation}

\begin{figure}[h]
\centering{}\includegraphics[width=1\columnwidth]{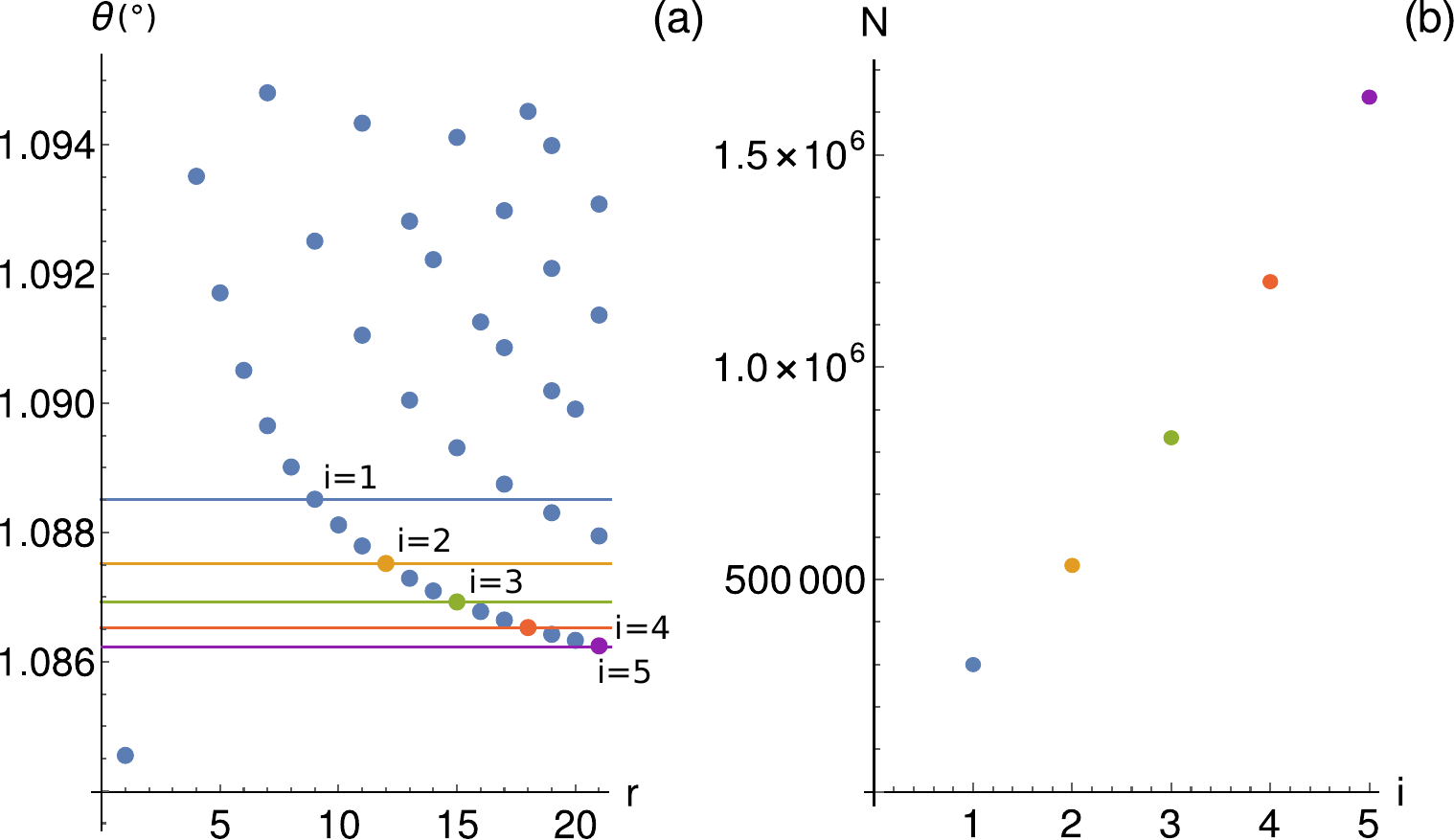}\caption{(a) Full set of commensurate structures that can be generated for
the range of angles $\theta\in[1.084^{\circ},1.095^{\circ}]$ and
$r\protect\leq21$. The horizontal lines indicate a possible choice
of a set of structures to make a finite-size scaling analysis. (b)
Total system size $N$ (number of sites) for the choice of structures
in (a). \label{fig:SUP_finite_size_scaling}}
\end{figure}

Our finite-size scaling procedure satisfies, for a set of $\mathcal{N}$
consecutive structure sizes, labelled by $i=1,...,\mathcal{N}$ and
characterized by $(m_{i},r_{i},n_{i})$ with $\theta_{i}=\theta(m_{i},r_{i})$
and $N_{i}=N(m_{i},r_{i},n_{i})$, such that:
\begin{itemize}
\item Every structure only contains one supercell ($n_{i}=1$);
\item $N_{i+1}>N_{i}$, with $N$ the total system size (number of sites);
\item $|\theta_{i+1}-\theta_{i}|/|\theta_{i}-\theta_{i-1}|<1$;
\item $|\theta_{2}-\theta_{1}|\sim10^{-5}{\rm rad}$.
\end{itemize}
In order to guarantee that the system's properties are not significantly
affected by differences in the angles used for structures in the finite-size
scaling analysis, we ensure that the variation in $\theta$ is sufficiently
small. We give an example of a possible set of structures in Fig.$\,$\ref{fig:SUP_finite_size_scaling}.
In Fig.$\,$\ref{fig:SUP_finite_size_scaling}(a), we plot all the
commensurate structures that can be generated for the range of angles
$\theta\in[1.084^{\circ},1.095^{\circ}]$ and $r\leq21$. The different
angles for a fixed $r$ were obtained by varying $m$. Out of these,
we chose the angles marked with horizontal lines for the finite-size
scaling analysis. With this choice, we can increase the system size
{[}Fig.$\,$\ref{fig:SUP_finite_size_scaling}(b){]}, and decrease
the required change in $\theta$ at the same time. The structures
in this example were used in the main text, in Fig.$\,$2(d--f).
The full variation in $\theta$ is below $10''$ (see Table$\,$\ref{tab:SUP_full_parameters_scalings}).

\section{Complementary information on exact diagonalization results}

\begin{figure}[h]
\centering{}\includegraphics[width=1\columnwidth]{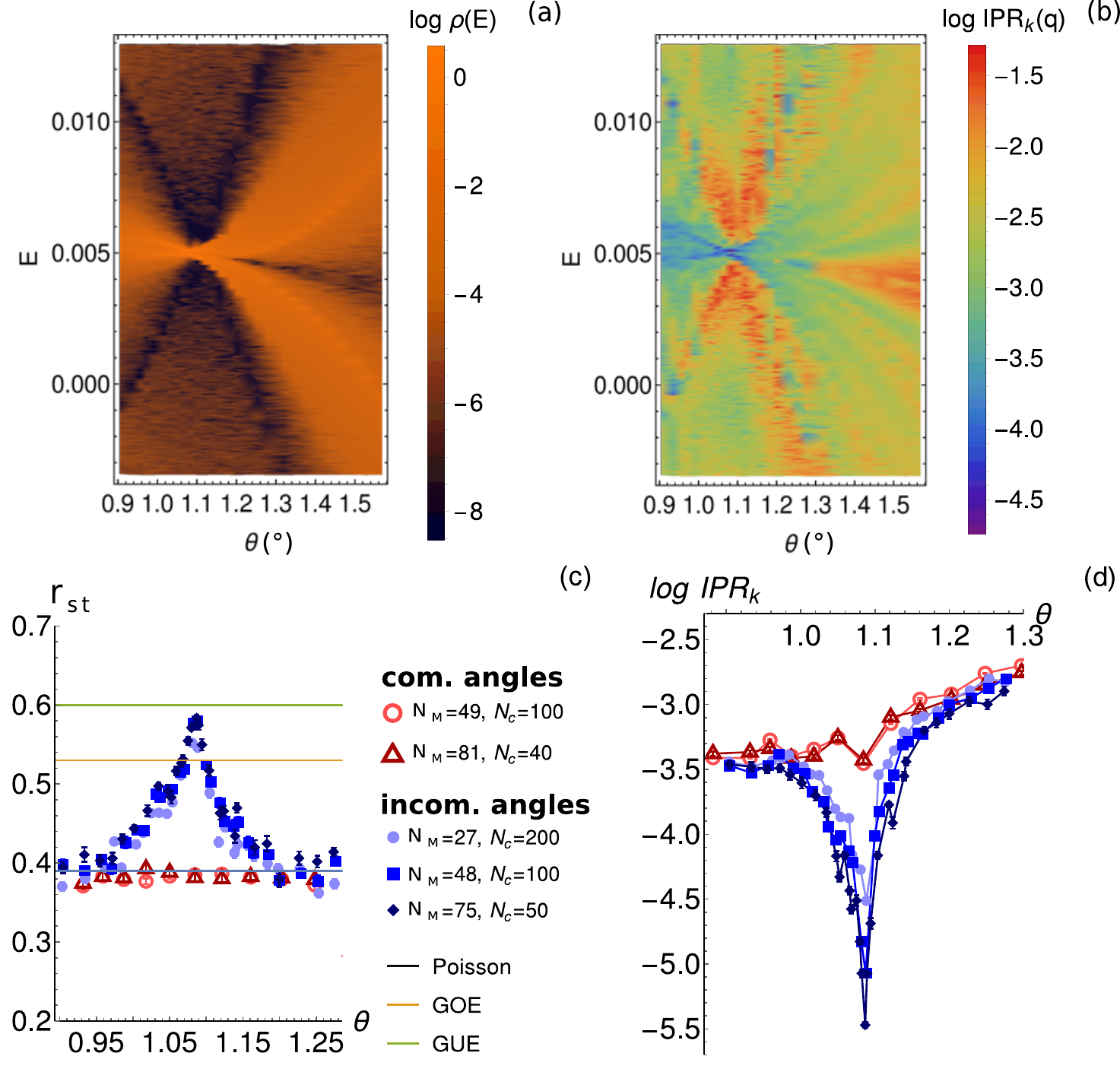}\caption{Analogous of Fig.$\,$1 of the main text. Panels~(a,b) are the same
as in Fig.$\,$1 (reproduced here for clarity), while the results
of panels$\,$(c,d) were instead obtained by following the VHS with
higher energy. $r_{{\rm st}}$ and ${\rm \mathcal{I}_{k}}$ were averaged
in a small energy around it. For different angles, we rescaled the
energy window used for averaging proportionally to the width of the
narrow-band, as described in the text. The energy is in units of $t$.\label{fig:SUP_fig1_largeDOS}}
\end{figure}

In Fig.$\,$1 of the main text, we plotted $\mathcal{I}_{k}$ and
$r_{{\rm st}}$ averaged over a fixed energy window containing the
narrow-band. Here we instead follow one of the VHS (the one of larger
energy), averaging over a small energy window $\delta E$ around it.
For the angle corresponding to the narrower band, $\delta E(\theta_{{\rm ft}})=2.5\times10^{-5}$.
For other angles, $\delta E(\theta)=\delta E(\theta_{{\rm ft}})\Delta E(\theta)/\Delta E(\theta_{{\rm ft}})$,
where $\Delta E(\theta)$ is the narrow-band's width for an angle
$\theta$. This accounts for the broadening of the DOS around the
VHS when the narrow-band becomes wider. The results are in Fig.$\,$\ref{fig:SUP_fig1_largeDOS},
where we can see that $r_{{\rm st}}(\theta)$ and $\mathcal{I}_{k}(\theta)$
become more peaked around $\theta=\theta_{{\rm ft}}$ than in Fig.$\,$1
of the main text, as expected.

\begin{figure}
\begin{centering}
\includegraphics[width=1\columnwidth]{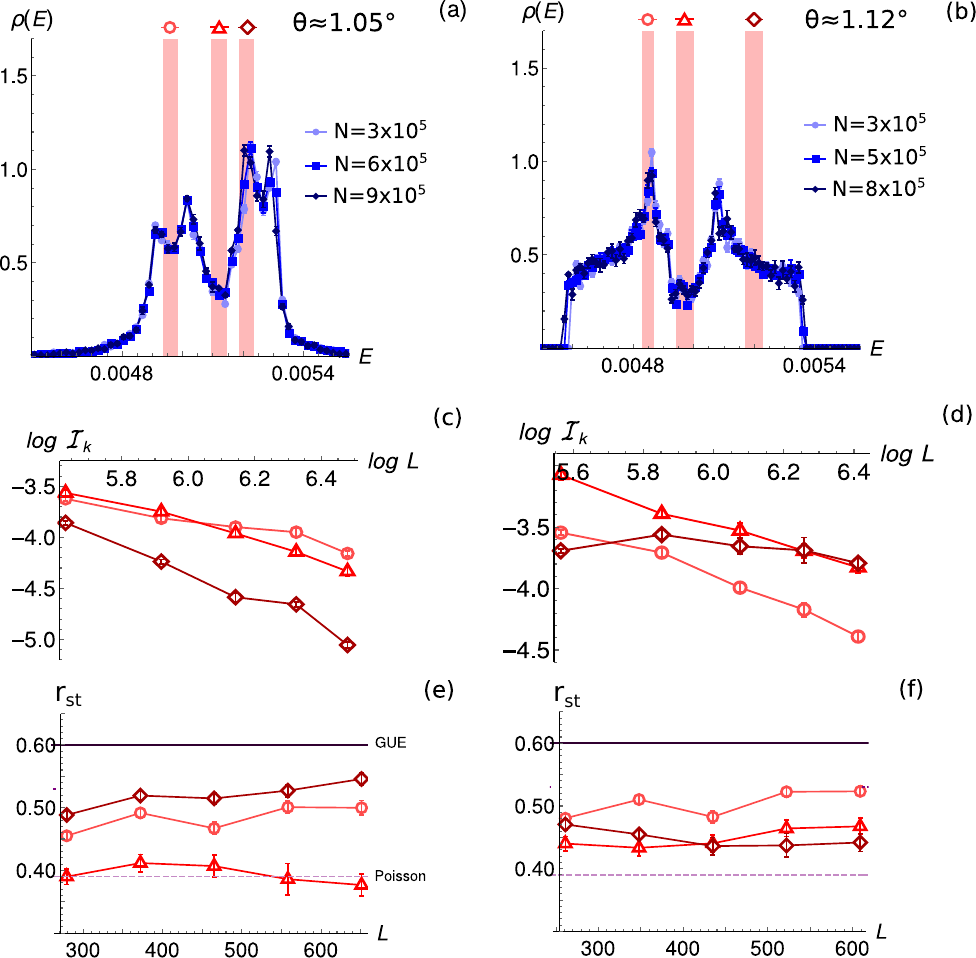}
\par
\end{centering}
\caption{(a,b) DOS for incommensurate structures of different sizes, for angles
slightly below (a) and above (b) the narrow-band regime. (c-f) Average
$r$ and $\mathcal{I}_{k}$ for a small energy window around selected
energies, depicted by the shaded areas in Figs.$\,$(a,b), as a function
of the linear system size $L\propto N^{1/2}$ (the largest systems
correspond to $N\gtrsim10^{6}$). Above the shaded areas we show the
plot markers used in the scaling analysis below the plots of $\rho(E)$
(in the same column). Complete information on the data used in Figs.$\,$(a-f)
can be found in table \ref{tab:SUP_full_parameters_scalings}. The energy is in units of $t$.\label{fig:SUP_fig2_complementary}}
\end{figure}

In Fig.$\,$2 of the main text, we analyzed sets of incommensurate
angles for which the bandwidth becomes very narrow. In Fig.$\,$\ref{fig:SUP_fig2_complementary}
we show a similar analysis for angles slightly above and below this
regime. Note that even though sub-ballistic behavior ($|\pd\log\mathcal{I}_{k}/\pd\log L|>0$
and $r_{{\rm st}}>r_{{\rm st}}^{{\rm Poisson}}$) is generically observed,
the large-$L$ limiting behavior is not always clear. For instance,
in Fig.$\,$\ref{fig:SUP_fig2_complementary}(f), we see that even
though $r_{{\rm st}}>r_{{\rm st}}^{{\rm Poisson}}$, it is not conclusive
whether $r_{{\rm st}}$ will converge to $r_{{\rm st}}^{{\rm GUE}}$
or $r_{{\rm st}}^{{\rm Poisson}}$ (or some intermediate value) upon
increasing $L$.

We finish this section by providing detailed information on the data
used in the plots of Fig.$\,$2 and Fig.$\,$3 of the main text and
Fig.$\,$\ref{fig:SUP_fig2_complementary}. Table~\ref{tab:SUP_full_parameters_scalings}
(\ref{tab:SUP_full_parameters_scalings-2}) details the parameters
used for calculations presented in Fig.$\,$2 of the main text and
Fig.$\,$\ref{fig:SUP_fig2_complementary} {[}Fig.$\,$3(c-e) of the
main text{]}.

\begin{table}
\begin{tabular}{|c|c|c|c|c|c|}
\hline 
 & $(m,r)$ & $\theta(\approx)$ & $N$ & $N_{{\rm M}}$ & $N_{c}$\tabularnewline
\hline 
\hline 
\multirow{5}{*}{Fig.$\,$2(a,c,e)} & (274,9) & $1.0690$ & 310 276 & 27 & 200\tabularnewline
\cline{2-6} \cline{3-6} \cline{4-6} \cline{5-6} \cline{6-6} 
 & (365,12) & $1.0699$ & 550 612 & 48 & 100\tabularnewline
\cline{2-6} \cline{3-6} \cline{4-6} \cline{5-6} \cline{6-6} 
 & (457,15) & $1.0682$ & 863 116 & 75 & 50\tabularnewline
\cline{2-6} \cline{3-6} \cline{4-6} \cline{5-6} \cline{6-6} 
 & (547,18) & $1.0709$ & 1 236 652 & 108 & 25\tabularnewline
\cline{2-6} \cline{3-6} \cline{4-6} \cline{5-6} \cline{6-6} 
 & (638,21) & $1.0712$ & 1 682 356 & 147 & 25\tabularnewline
\hline 
\multirow{5}{*}{Fig.$\,$2(b,d,f)} & (269,9) & $1.0885$ & 299 236 & 27 & 200\tabularnewline
\cline{2-6} \cline{3-6} \cline{4-6} \cline{5-6} \cline{6-6} 
 & (359,12) & $1.0875$ & 532 948 & 48 & 100\tabularnewline
\cline{2-6} \cline{3-6} \cline{4-6} \cline{5-6} \cline{6-6} 
 & (449,15) & $1.0869$ & 833 644 & 75 & 50\tabularnewline
\cline{2-6} \cline{3-6} \cline{4-6} \cline{5-6} \cline{6-6} 
 & (539,18) & $1.0865$ & 1 201 324 & 108 & 25\tabularnewline
\cline{2-6} \cline{3-6} \cline{4-6} \cline{5-6} \cline{6-6} 
 & (629,21) & $1.0862$ & 1 635 988 & 147 & 25\tabularnewline
\hline 
\multirow{5}{*}{Fig.$\,$\ref{fig:SUP_fig2_complementary}(a,c,e)} & (280,9) & $1.0464$ & 323 788 & 27 & 200\tabularnewline
\cline{2-6} \cline{3-6} \cline{4-6} \cline{5-6} \cline{6-6} 
 & (373,12) & $1.0474$ & 574 612 & 48 & 100\tabularnewline
\cline{2-6} \cline{3-6} \cline{4-6} \cline{5-6} \cline{6-6} 
 & (466,15) & $1.0479$ & 896 884 & 75 & 50\tabularnewline
\cline{2-6} \cline{3-6} \cline{4-6} \cline{5-6} \cline{6-6} 
 & (559,18) & $1.0483$ & 1 290 604 & 108 & 25\tabularnewline
\cline{2-6} \cline{3-6} \cline{4-6} \cline{5-6} \cline{6-6} 
 & (652,21) & $1.0485$ & 1 755 772 & 147 & 25\tabularnewline
\hline 
\multirow{5}{*}{Fig.$\,$\ref{fig:SUP_fig2_complementary}(b,d,f)} & (262,9) & $1.1171$ & 284 116 & 27 & 200\tabularnewline
\cline{2-6} \cline{3-6} \cline{4-6} \cline{5-6} \cline{6-6} 
 & (349,12) & $1.1182$ & 504 148 & 48 & 100\tabularnewline
\cline{2-6} \cline{3-6} \cline{4-6} \cline{5-6} \cline{6-6} 
 & (436,15) & $1.1188$ & 786 844 & 75 & 50\tabularnewline
\cline{2-6} \cline{3-6} \cline{4-6} \cline{5-6} \cline{6-6} 
 & (523,18) & $1.1192$ & 1 132 204 & 108 & 25\tabularnewline
\cline{2-6} \cline{3-6} \cline{4-6} \cline{5-6} \cline{6-6} 
 & (610,21) & $1.1195$ & 1 540 228 & 147 & 25\tabularnewline
\hline 
\end{tabular}

\caption{Complete information on data used in Fig.$\,$2 of the main text and
Fig.$\,$\ref{fig:SUP_fig2_complementary}. In the first column we
only indicate $(m,r)$ because $n=1$ (all structures are incommensurate).
$\theta$ is the twist angle, $N$ is the total number of sites in
the system, $N_{{\rm M}}$ is the number of moiré patterns in the
system and $N_{c}$ is the total number of realizations (including
averages over random twisted boundary conditions and shifts $\bm{\delta}_{{\rm t}}$).
Each row corresponds to a point in Figs.$\,$2(d,f) and Figs.$\,$\ref{fig:SUP_fig2_complementary}(d,f).
Note that in Fig.$\,$2(a) and Fig.$\,$\ref{fig:SUP_fig2_complementary}(a),
the DOS is plotted for the three smallest system sizes used (first
three rows in the corresponding section of the table). \label{tab:SUP_full_parameters_scalings}}
\end{table}

\begin{table}
\begin{tabular}{|c|c|c|c|c|c|}
\hline 
 & $(m,r,n)$ & $\theta(\approx)$ & $N$ & $N_{{\rm M}}$ & $N_{c}$\tabularnewline
\hline 
\hline 
\multirow{4}{*}{Fig.$\,$3(c-e),com.} & (7,1,25) & \multirow{4}{*}{$4.4085$} & 422 500 & 625 & 35\tabularnewline
\cline{2-2} \cline{4-6} \cline{5-6} \cline{6-6} 
 & (7,1,30) &  & 608 400 & 900 & 24\tabularnewline
\cline{2-2} \cline{4-6} \cline{5-6} \cline{6-6} 
 & (7,1,35) &  & 828 100 & 1225 & 18\tabularnewline
\cline{2-2} \cline{4-6} \cline{5-6} \cline{6-6} 
 & (7,1,40) &  & 1 081 600 & 1600 & 13\tabularnewline
\hline 
\multirow{7}{*}{Fig.$\,$3(c-e),incom.} & (190,27,1) & $4.3868$ & 165 892 & 243 & 102\tabularnewline
\cline{2-6} \cline{3-6} \cline{4-6} \cline{5-6} \cline{6-6} 
 & (232,33,1) & $4.3907$ & 247 372 & 363 & 68\tabularnewline
\cline{2-6} \cline{3-6} \cline{4-6} \cline{5-6} \cline{6-6} 
 & (274,39,1) & $4.3934$ & 345 076 & 507 & 49\tabularnewline
\cline{2-6} \cline{3-6} \cline{4-6} \cline{5-6} \cline{6-6} 
 & (316,45,1) & $4.3954$ & 459 004 & 675 & 36\tabularnewline
\cline{2-6} \cline{3-6} \cline{4-6} \cline{5-6} \cline{6-6} 
 & (358,51,1) & $4.3970$ & 589 156 & 867 & 28\tabularnewline
\cline{2-6} \cline{3-6} \cline{4-6} \cline{5-6} \cline{6-6} 
 & (400,57,1) & $4.3982$ & 735 532 & 1083 & 23\tabularnewline
\cline{2-6} \cline{3-6} \cline{4-6} \cline{5-6} \cline{6-6} 
 & (442,63,1) & $4.3992$ & 898 132 & 1323 & 18\tabularnewline
\hline 
\end{tabular}

\caption{Complete information on data used in Figs.$\,$3(c-e) of the main
text, for the data points corresponding to commensurate and incommensurate
structures. Each row corresponds to a point in Figs.$\,$3(d,e). Note
that in Fig.$\,$3(c), the DOS is plotted for the smallest commensurate
and incommensurate system size used (first row in the corresponding
section of the table). See table \ref{tab:SUP_full_parameters_scalings}
for description of the parameters. \label{tab:SUP_full_parameters_scalings-2}}
\end{table}

\section{Band inversion}

Fig.$\,$1(a) of the main text raises the question on whether there
is a band inversion as a function of $\theta$.

To answer that, we can, similarly to Ref.$\,$\cite{Pixley2018},
define projectors in both bands for a larger angle and project them
into new states obtained upon decreasing the angle. The results evaluate
the subspace with which the new states have the larger overlap. However,
to use this method, the system sizes for different angles should be
fixed. This is not achievable with our method of building the system.

An alternative way of observing the same angle-driven physics is by
varying the interlayer coupling. Increasing the interlayer coupling
for an angle above the narrow-band regime leads to similar results as
decreasing the angle. The advantage is that the system size can be
made fixed in the former.

We generalize the interlayer coupling $t_{\perp}(r)$ of the main
text, defining a parameter $V$ as

\begin{equation}
\begin{aligned}t_{\perp}(V,r)=V\Bigg(\frac{d_{\perp}^{2}}{d_{\perp}^{2}+r^{2}}V_{pp\sigma}\left(\sqrt{d_{\perp}^{2}+r^{2}}\right)\\
+\frac{r^{2}}{d_{\perp}^{2}+r^{2}}V_{pp\pi}\left(\sqrt{d_{\perp}^{2}+r^{2}}\right)\Bigg).
\end{aligned}
\end{equation}

We vary $V$ for a fixed commensurate (incommensurate) angle $\theta\approx1.2972^{\circ}$
($\theta\approx1.2930^{\circ}$) corresponding to the structure $(m,r)=(25,1)$
{[}$(m,r)=(301,12)${]}. For $V=1$ we define the projector into the
lowest and highest energy bands, respectively $B_{+}$ and $B_{-}$
as

\begin{equation}
P_{\pm}(V=1)=\sum_{v\in B_{\pm}}\ket v\bra v,
\end{equation}
where $\left\{ \ket v\right\} $ is the set of eigenstates obtained
through exact diagonalization.

We then vary $V$ and for each eigenstate $\ket{v(V,E)}$ in the energy
window of interest we define the quantity

\begin{equation}
\mathcal{C}=\bra{v(V,E)}\left[P_{+}(V=1)-P_{-}(V=1)\right]\ket{v(V,E)}\label{eq:projection_C}
\end{equation}

\noindent which should be $\mathcal{C}=1$ if $\ket{v(V,E)}$ belongs
to $B_{+}$, $\mathcal{C}=-1$ if it belongs to $B_{-}$ and $\mathcal{C}=0$
if it is orthogonal to both.

\begin{figure}[H]
\centering{}\includegraphics[width=1\columnwidth]{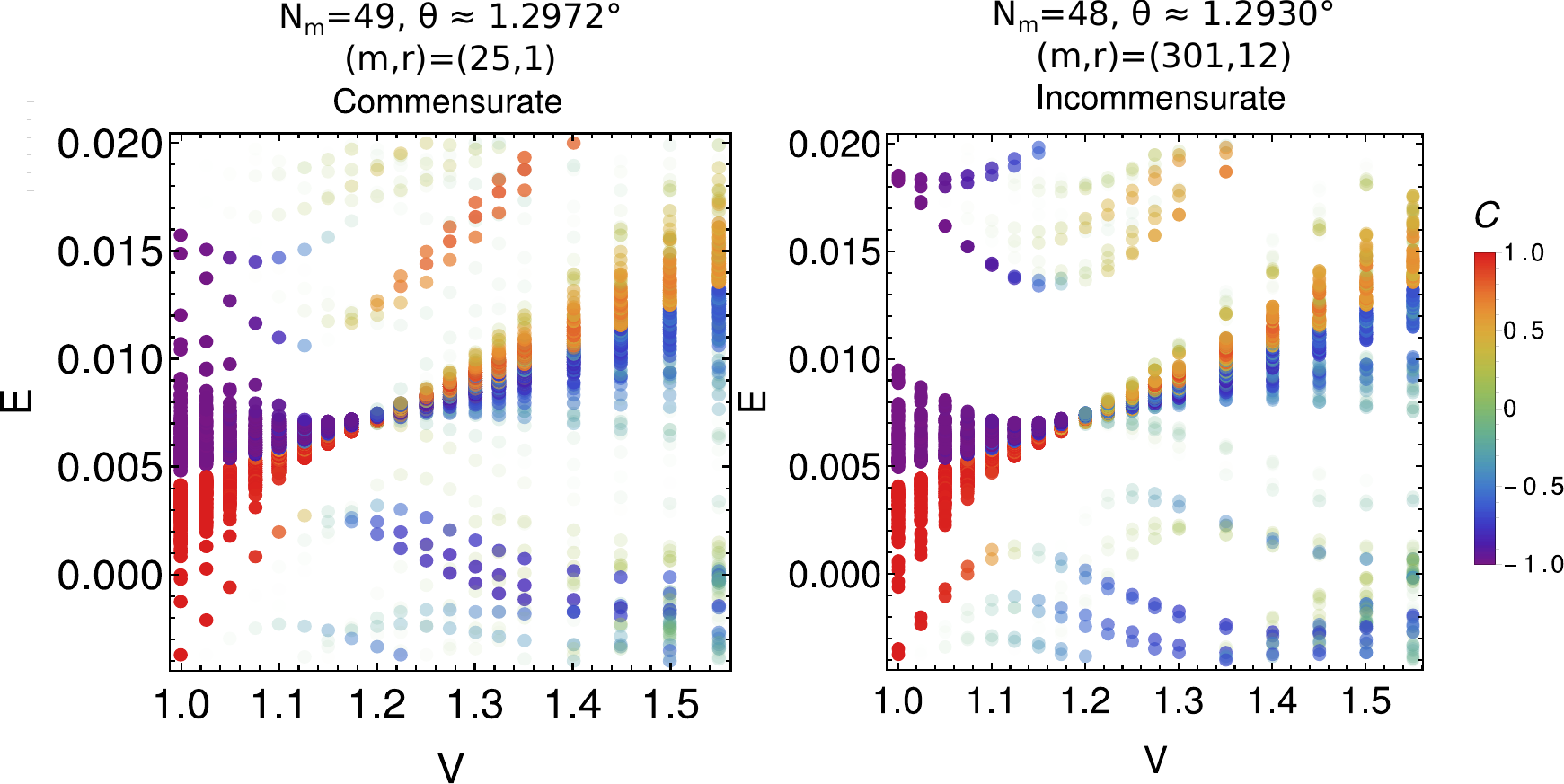}\caption{Quantity $\mathcal{C}$ defined in Eq.$\,$\ref{eq:projection_C}
for every eigenstate within the energy window $E\in[0,0.02]$, for
variable $V$. The results were obtained for a fixed random twist
and stacking, for the commensurate (incommensurate) structures $(m,r)=(25,1)$
{[}$(m,r)=(301,12)${]}. The system sizes were chosen to contain $N_{m}=48$
($N_{m}=49$) moiré patterns. The energy is in units of $t$. \label{fig:com_vs_incom_projector}}
\end{figure}

The results are in Fig.$\,$\ref{fig:com_vs_incom_projector} for
fixed configurations in the commensurate and incommensurate cases.
In both cases, the results suggest that a band inversion occurs when
the narrow-band regime is reached. The higher (lower) energy states
for larger $V$ (or equivalently, smaller $\theta$) have a larger
projection in $B_{-}$ ($B_{+}$).

\newpage

\section{Robustness of incommensurability effects to disorder}

\begin{figure}[h]
\begin{centering}
\includegraphics[width=1\columnwidth]{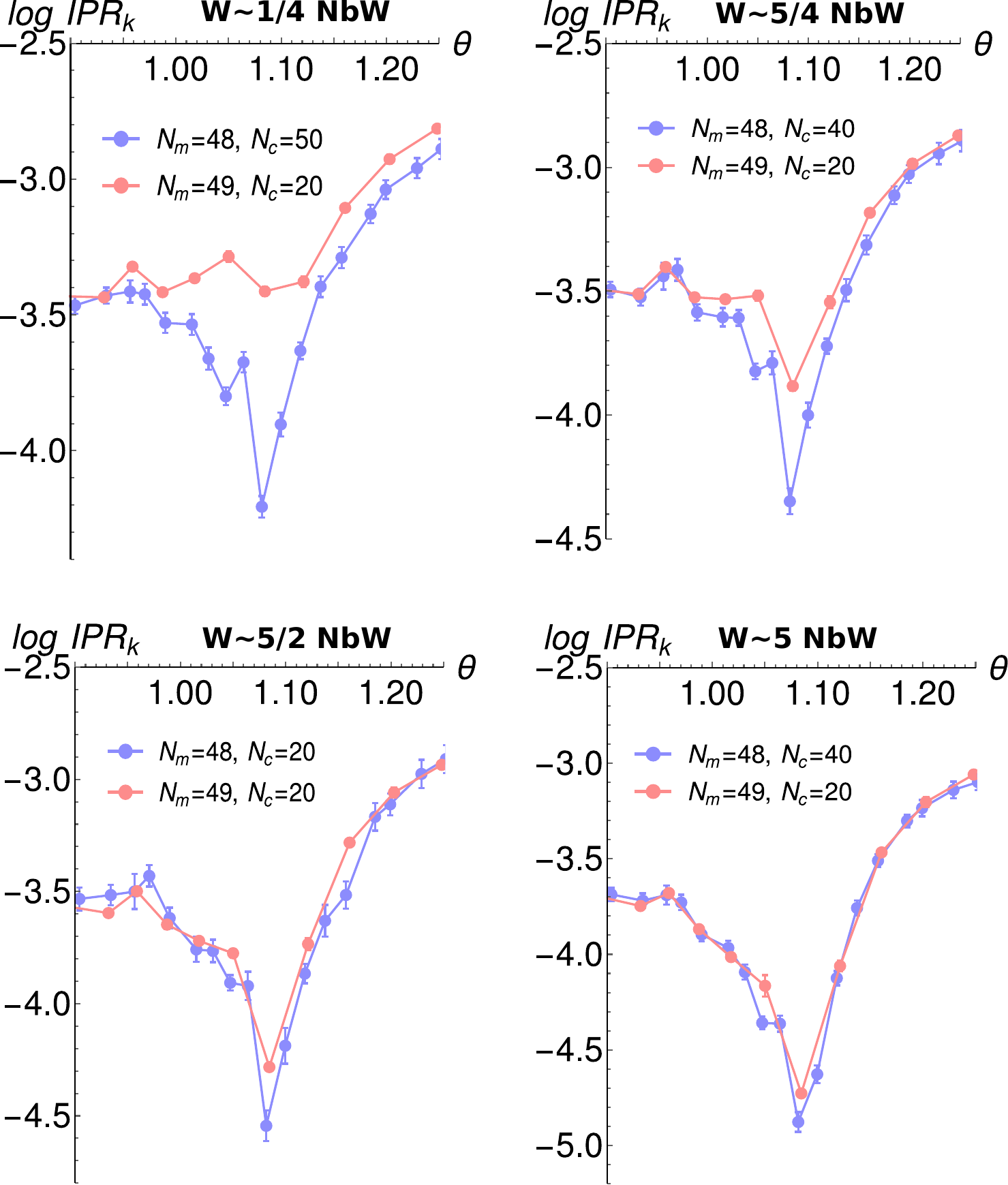}
\par\end{centering}
\caption{ $\mathcal{I}_{k}$ for commensurate (red points) and incommensurate (blue points) angles and variable disorder strength $W$. $NbW$ is the smallest observed bandwidth, $\Delta E \approx 0.0004$, at $\theta = \theta_{\textrm{ft}}\approx 1.09^{\circ}$. The results were obtained for a fixed number of Moiré patterns, $N_M=48$ ($N_M=49$) for incommensurate (commensurate) angles. \label{iprk_disorder_sup}}
\end{figure}

\begin{figure}[h]
\begin{centering}
\includegraphics[width=1\columnwidth]{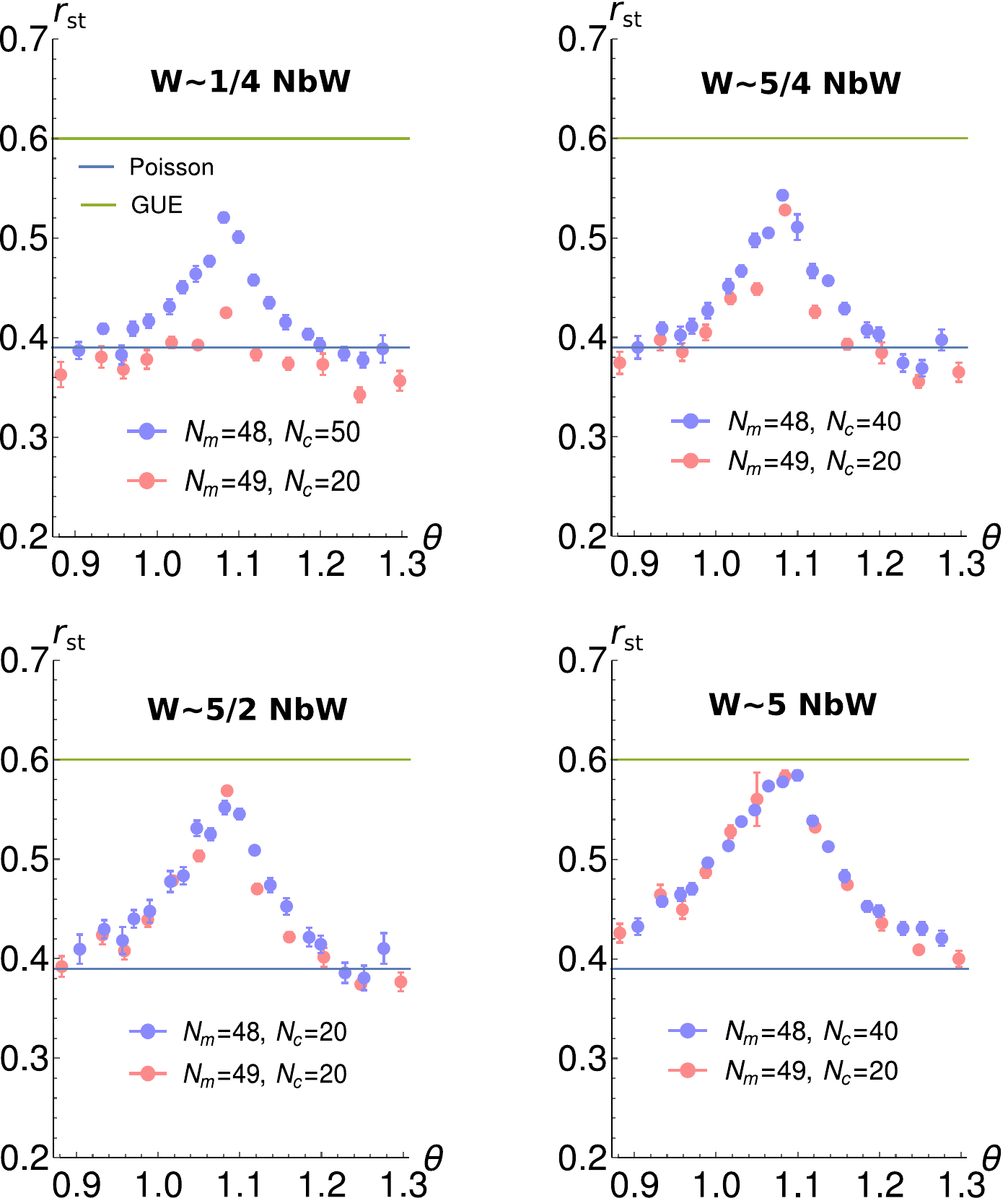}
\par\end{centering}
\caption{ $r_{\textrm{st}}$ for commensurate (red points) and incommensurate (blue points) angles and variable disorder strength $W$. $NbW$ is the smallest observed bandwidth, $\Delta E \approx 0.0004$, at $\theta = \theta_{\textrm{ft}}\approx 1.09^{\circ}$. The results were obtained for a fixed number of Moiré patterns, $N_M=48$ ($N_M=49$) for incommensurate (commensurate) angles. \label{lss_disorder_sup}}
\end{figure}

In this section, we study how the effects presented in the main text are affected by the presence of disorder. In particular, we use disorder of the Anderson type, consisting of random on-site energies $\epsilon_i$ sampled according to a box distribution:

\begin{equation}
P_{W}(\epsilon_{i})=\frac{1}{W}\Theta\left(\frac{W}{2}-\left|\epsilon_{i}\right|\right),
\end{equation}

\noindent where $W$ is the disorder strength and $\Theta(x)$ is the Heaviside function. We carried out an identical study to the one made in Fig.$\,$1 of the main text for the clean case. In particular, we averaged $\mathcal{I}_{k}$ and $r_{\textrm{st}}$ for the states within the same energy window used there ($E\in[0.004,0.006]$). The results for different $W$ are in Figs.$\,$\ref{iprk_disorder_sup},\ref{lss_disorder_sup}. We studied systems with $N_M=48$ ($N_M=49$) Moiré patterns for incommensurate (commensurate) angles. Similar system sizes were used for commensurate and incommensurate angles so that a fair comparison could be made. The reason follows: it is well known that any finite amount of disorder in 2D gives rise to real-space localization \cite{PhysRevLett.42.673}. However, for the disorder strengths we are concerned with, the localization length is unrealistically large, much larger than the system sizes that we are studying and than the sample sizes that are studied experimentally. Still, different localization regimes may be reached for different system sizes and a fair comparison between commensurate and incommensurate strucutures should be made for similar system sizes. 

The results in Figs.$\,$\ref{iprk_disorder_sup},\ref{lss_disorder_sup} indicate that the differences observed between commensurate and incommensurate angles in the clean case are robust up to a relatively strong disorder, larger than the narrow-band's width (we define this width, $\textrm{NbW}$, as the smallest observed bandwidth, at $\theta = \theta_{\textrm{ft}}\approx 1.09^{\circ}$). For stronger disorder strengths, the results obtained for commensurate angles approach the results obtained for incommensurate angles. 

The results in this section have two important implications. Firstly, they show that the effects of incommensurability are comparable to the effects of a strong disorder, reinforcing their importance. Secondly, they provide yet another indication of the fragility of the results obtained for commensurate structures, that are much less robust to disorder than the results obtained for incommensurate structures. Both points reinforce the importance of incommensurate structures to the physics of tBLG. 


\section{Technical details on the computation of $\mathcal{I}_{k}$}

\begin{figure}[h]
\begin{centering}
\includegraphics[width=1\columnwidth]{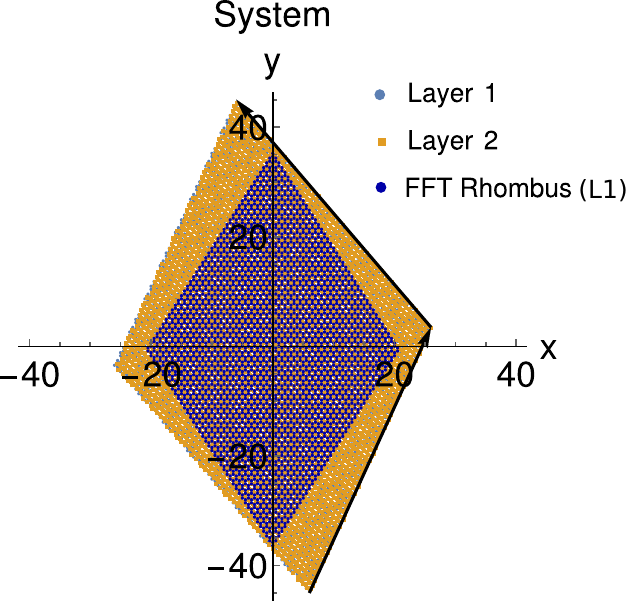}
\par\end{centering}
\caption{Example structure $(m,r,n)=(5,3,8)$. The arrows correspond to $8\bm{v}_{1}$
and $8\bm{v}_{2}$, where $\bm{v}_{1},\bm{v}_{2}$ are the superlattice
vectors of the supercell defined by $(m,r)=(5,3)$. In darker blue
we represent the smaller rhombus selected in the unrotated layer (layer
1) for which we apply a Fast Fourier Transform in order to compute
$\mathcal{I}_{k}$. \label{fig:SUP_fft_rhombus}}
\end{figure}

We made use of the Fast Fourier Transform (FFT) method to compute
the wavefunction in the momentum-space. We do this only for the unrotated
layer (layer 1) without loss of generality: there is no ``preferred'' layer in our model and
therefore the wavefunction is similar in both.

We cut a rhombus of linear size $L\propto N^{1/2}$ along the directions
of the lattice vectors of the honeycomb lattice. $L$ sets the momentum-space
resolution for computing $\mathcal{I}_{k}$. In the main text, we
always use the definition of $L$ given here, for consistency. The
total number of sites $N$ is slightly larger than $4L^{2}$. Nonetheless,
$L$ can be seen as the total linear system size for all the discussions
in the main text.

After cutting the smaller rhombus, we apply the FFT to the real-space
wavefunction in that rhombus and finally compute $\mathcal{I}_{k}$
using the resulting momentum-space wavefunction.

An example of this procedure is in Fig.$\,$\ref{fig:SUP_fft_rhombus}
for a structure with $(m,r,n)=(5,3,8)$. We notice that the ratio
between the rhombus size and total system size becomes larger when
we use structures with $\mod(r,3)=0$. For this reason we use structures
of this type in the main text, but the obtained results are also valid
for incommensurate structures with $\mod(r,3)\neq0$.

\section{Technical details on conductance calculations and additional results}

\begin{figure}[h]
\begin{centering}
\includegraphics[width=1\columnwidth]{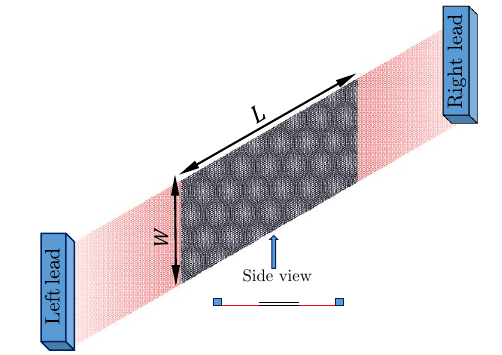}
\par\end{centering}
\caption{\label{fig:Conductance_setup}Two-terminal conductance setup used
in this work. The scattering region, in black, is tBLG of parallelogram
shape. The leads are semi-infinite finite-width graphene ribbons.
The side-view of the setup depicts that the tBLG region is connected
to the leads through the bottom layer.}
\end{figure}

Conductance calculations were perfomed for the system depicted in
Fig.~\ref{fig:Conductance_setup}: a parallelogramical tBLG scattering
region, with a side of $W$ and a base of $L$, with semi-infinite
monolayer graphene leads attached to the bottom layer of the tBLG.
In order to avoid the overlap in energy of the leads Dirac points
(with a vanishing DOS) with the tBLG narrow-bands, we considered doped
leads, with the left(right) lead shifted in energy by positive(negative)
onsite potentials of $0.15t$, with respect to the tBLG region. The
conductance, at zero temperature and in the linear regime, was computed
within the Landauer approach. The conductance per width is given by
$G=G_{0}T\left(\epsilon\right)/W$, where $G_{0}=2e^{2}/h$ is the
conductance quantum and $T\left(\epsilon\right)$ is the transmission
at energy $\epsilon$. The transmission was computed using the Kwant
package \cite{Kwant_paper}. The structures simulated have a width
of $W=1200a$, where $a$ is the lattice constant of graphene, and
contain from $0.8\times10^{6}$ up to $6.2\times10^{6}$ carbon atoms
for increasing lengths, $L$, of the tBLG region. For each system
size, $G$ was averaged over 25 different stackings in tBLG region,
where the center of rotation of the top layer is shifted with respect
to the bottom layer along the vectors $\bm{\delta}_{t}=x\mathbf{a}_{1}+y\mathbf{a}_{2}$,
with $x,y\in\left\{ 0,1/4,2/4,3/4,1\right\} $ and $\mathbf{a}_{1/2}$
the primitive vectors of the top layer.

\begin{figure}
\centering{}\includegraphics[width=1\columnwidth]{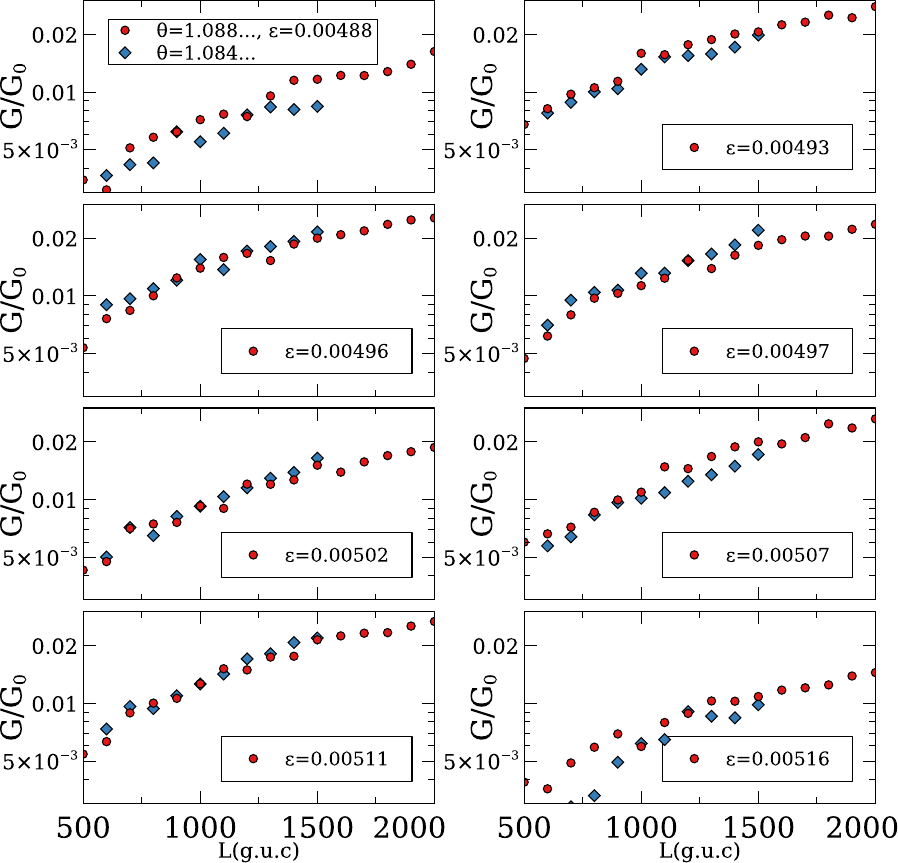}\caption{Conductance results for $t_{\perp}=0.48{\rm eV}$ in the narrow-band
regime, for two close commensurate and incommensurate angles. Commensurate angle: $\theta \approx 1.0845, (m,r)=(30,1)$ (blue diamonds); Incommensurate angle: $\theta \approx 1.0888, (m,r)=(29881,1000)$ (red circles). This angle has a unit cell with approximately $10^{10}$ sites, much larger than the simulated system sizes. The energy is in units of $t$. \label{fig:Conductance-realistic}}
\end{figure}

For the realistic tight binding model considered in the main text,
where $t_{\perp}=0.48\,{\rm eV}$, we computed $G$ for both commensurate
and incommensurate angles in the narrow-band regime, obtaining very
similar results (see Fig.$\,$\ref{fig:Conductance-realistic}). Furthermore,
$G(L)$ increases with $L$ up to the maximal system size we were
able to simulate. Note that for ballistic transport we expect an $L$-independent
conductance, whereas $G(L)\propto L^{-1}$ for diffusion. Therefore,
an increasing $G(L)$ with $L$ implies that finite size effects are
too severe to draw any conclusions. Such finite-size effects are not
surprising as even the largest systems simulated contain only\textcolor{black}{{}
$N_{\t M}\sim10^{2}$} moiré pattern cells.

To reduce finite-size effects, we considered a model with an increased
interlayer coupling of $t_{\perp}=1.9\,{\rm eV}$. For this increased
interlayer hopping, the narrow-bands occur for $\theta_{\text{ft}}\sim4.4^{\circ}$.
Thus, the maximum size numerically attainable will include significantly
more moiré pattern cells and finite-size effects are reduced. For
this modified model, the number of moiré pattern cells in the scattering
region ranges from $\gtrsim2.8\times10^{3}$, for the smallest system
size, up to $\gtrsim2.1\times10^{4}$, for the largest system size
simulated. $G(L)$ was again computed for two close commensurate and
incommensurate angles in the new narrow-band regime, $\theta_{\text{ft}}\sim4.4^{\circ}$,
with the results shown in Fig.$\,$3(a,b) of the main text.

\begin{figure*}
\centering{}\includegraphics[width=1\textwidth]{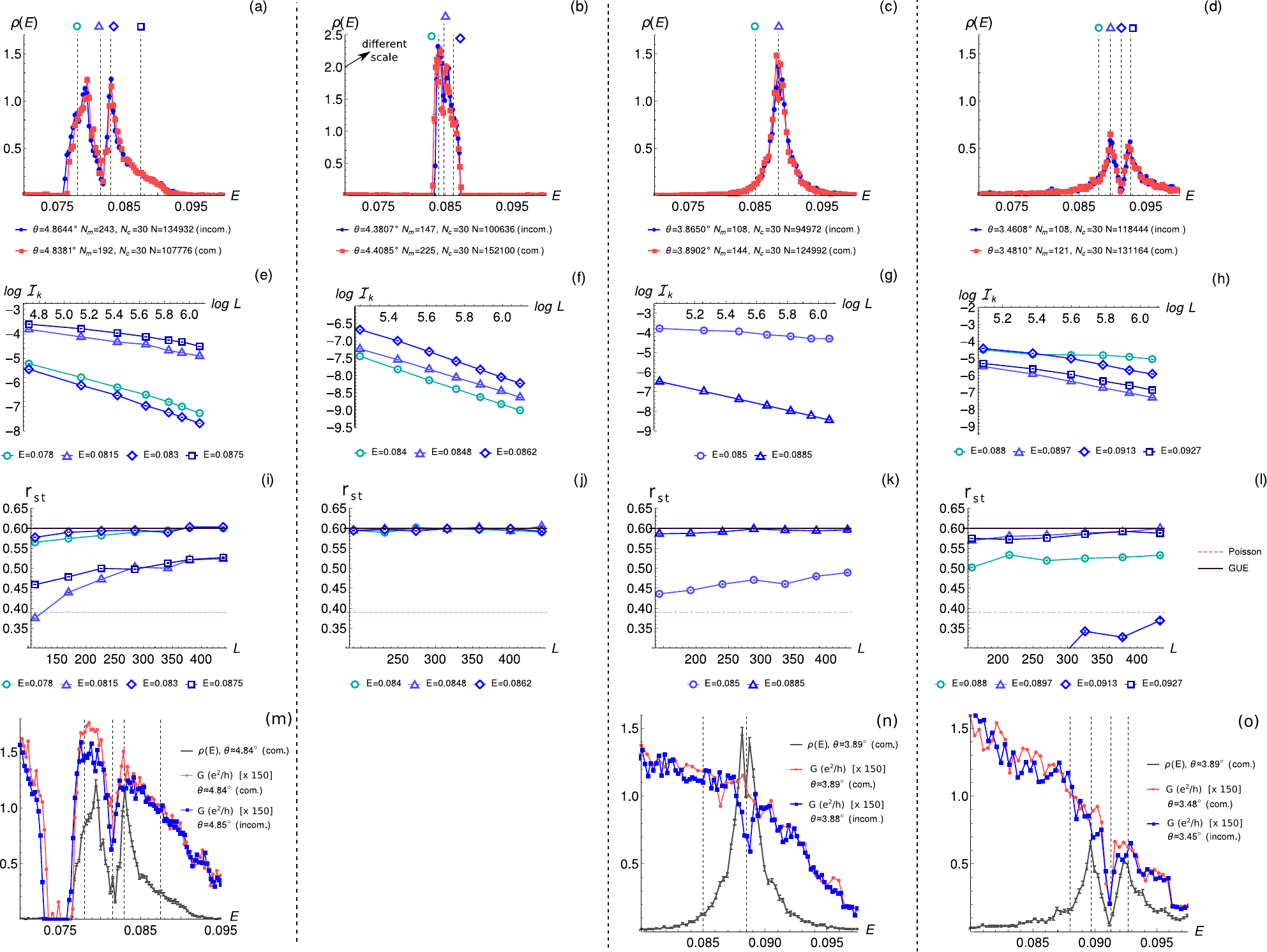}\caption{(a-d) DOS for commensurate (full red squares) and incommensurate (full
blue circles) structures around the narrow-band regime. For each column,
we present results for two close commensurate and incommensurate angles.
$N_{{\rm M}}$, $N_{c}$, and $N$ are respectively the number of
moiré pattern cells, configurations and total number of sites in the
system. (e-l) We make a finite-size scaling analysis for a set of
incommensurate angles close to the angles depicted in (a-d), and plot
the average $\mathcal{I}_{k}$ {[}panels$\,$(e-h){]} and $r_{{\rm st}}$
{[}panels$\,$(i-l){]} below the respective figure in the first row
{[}panels$\,$(a-d){]}. The average is taken over a small energy window
around some selected energies, depicted by the dashed lines in panels.$\,$(a-d).
Above these dashed lines we show the plotmarkers used in the corresponding
scaling analysis in panels$\,$(e-l). Note that the scalings in each
column are relative to the upper row figure in the same column. (m-o)
Conductance, $G(e^{2}/h)$, as a function of energy for the commensurate
angles used in the upper panels and for incommensurate angles close
to them. The system sizes were fixed to $L=3000$. To simulate the
incommensurate angles, structures with a unit cell much larger than
the system size were considered. The DOS for the commensurate angles
is also shown in gray {[}for the incommensurate angles, the results
are very similar, as shown in panels (a,b,d){]}. The energy is in units of $t$. \textcolor{red}{\label{fig:Conductance-energy-plusED}}}
\end{figure*}

To complement the conductance results in the main text, we show in
Fig.$\,$\ref{fig:Conductance-energy-plusED} the conductance as a
function of energy for additional commensurate and incommensurate
twist angles close to the narrow-band regime, together with an ED
analysis. With the ED analysis (first three rows) we show that similarly
to the case of a smaller $t_{\perp}$, different scaling behaviours
can be observed for different energies within the narrow-band regime.
In the last row, we show the conductance as a function of energy for
a fixed large system size and for commensurate and incommensurate
angles close to the ones used in the ED analysis. We see that the
conductance is smaller for an incommensurate angle in comparison with
a nearby commensurate angle around energies for which the $\mathcal{I}_{k}$
scales faster with system size and $r_{{\rm st}}$ is closer to $r_{{\rm GUE}}$.
For such energies, the conductance for incommensurate angles is expected
to decrease even more if larger systems are considered.

\end{document}